# Mechanism of the Fano resonance in a planar metamaterials: Analysis from the coupled two-oscillator model

Kum-Song Song and Song-Jin Im

Department of Physics, **Kim Il Sung** University, Pyongyang, DPR Korea

**Abstract** Mechanism of the Fano resonances in planar metamaterials demonstrate based on the coupled two-oscillator model. We have described the optical spectrums like reflectance and transmittance near the resonances of bright mode (continuum mode) and dark mode (discrete mode) and explained their optical properties by the Fano formulism. the Fano formulism of the resonances in the planar metamaterials can predict the asymmetric shape line and radiative properties occurring in reflectance and transmittance from the coupling between bright and dark modes.

## Introduction

The coupled two-oscillator model is very rational to describe the Fano resonances due to the hybridization of bright mode and dark mode in metamaterials or the coupling of molecular vibrational excitations with broadband plasmon resonances, which act as a continuum state and a discrete state[1-4].

So far the coupled two oscillator model in these kind of systems have been introduced to explain the physical behavior of the Fano-like spectrum phenomenologically[5,6].

One of significances from the Fano formulism in physics is capable to clear and simply explain many physical phenomenon revealing asymmetric spectrum, based on a few and simple factors as the reduced energy and asymmetric shape factor so on. The coupled two-oscillator model is very simple and useful to explain the physical behaviors of resonant phenomenon occurring in metamaterials and plasmonics[7]. So far, this model has used a tool to interpret the classical analog of electromagnetically induced transparency (EIT) or absorption (EIA) experimentally measured in metamaterials and plasmonics [8,9,18-23,40,41], but didn't explain in detail the fundamental factor of model used in experiment. for example, the relationship between the geometry of dipole (bright) mode and quadruple (dark) mode as size or gap with macro-experimental data like as reflectance or transmittance in dolmen structure [8,43].

It is developed new formulism, which can describe the relation between macro response and micro behavior of bright atom (dipole mode) and dark atom(quadruple) in single-layer planar metamaterials using radiating two-oscillator model[1,32,33,42]. When bright atoms are coupled with dark atoms and affected by the external field in metamaterials and plasmonics, the interaction between them often leads to the spectrum such as the electromagnetically induced transparency or absorption in atom physics[13-15], more generally appear as the Fano-like spectrum[24-30, 35-39].

It also developed one method that can derive the Fano like spectrum from the classical model of two coupled oscillators [25-28]. Based on this method, they calculated the reflectance from individual resonant positions and damping factors in the system composed of a bright atom and dark atom [27, 37].

In this paper, based on the idea that derive reflectance, transparency and absorption from the surface conductivity [1,42] and the Fano formulism of classical two coupled oscillators [4,7,38], we have a mind to reveal the relationship between micro properties and macro optical properties in real systems.
Through such study, it will give us a key that reveals the nature of EIT and EIA in metamaterials, for example, in easy fabricated single-layer metamaterials (or meta-surface)

The results of this paper will give a useful help to design and develop a variable of functional devices such as a high sensitive-sensor [19, 34, 38], slow light application [17, 18, 38], light switching so on[31].

## 1. The derive of the Fano formulism in macro-optical properties from two coupled oscillator model

The interaction of a bright atom coupled a dark atom with external field can better be described by two coupled oscillator model. The study of such systems mostly undergo in fabrication simple planar metamaterials (or meta-surface), typical example is the dolmen structure.

The dolmen structure is comprised of a metal rod (dipole or bright mode) exhibiting dipole radiation and two parallel metal rods (quadruple or dark mode) perpendicular to one, the bright mode interact directly with external field, but the dark mode interact indirectly with external field through the excitation the bright one (for example, the magnetic field occurring from bight mode).

The quadruple mode may be affected by the dipole one of neighborhood cell in-plane dolmen structure, but the stacked dolmen structure is comprised of almost closed system and experience little effect of period, since dipole mode stack on top of quadruple.

These electrical dipole-quadruple systems can be described in terms of RLC model coupled by a shunt capacitor [20], written in terms of two coupled oscillator model as follows:

$$L_1 \ddot{p}(t) + R_1 \dot{p}(t) + \frac{1}{C_1} p(t) + \frac{1}{C_c}[p(t) - q(t)] = U_{ext}$$
$$L_2 \ddot{q}(t) + R_2 \dot{q}(t) + \frac{1}{C_1} q(t) + \frac{1}{C_c}[q(t) - p(t)] = U_{int} = M\ddot{p}(t)$$
(1)

Where $p(t)$ is the charge of dipole mode and $q(t)$ is the one of quadruple. $M$ is the mutual induction of quadruple mode induced by magnetic field occurring from dipole mode, $U_{ext}$ is the electromotive force induced by the external field and $U_{int}$ is the electromotive force induced by the dipole mode, proportional to the current of dipole mode.

Assuming the harmonic time approach, the equation follows as:

$$-\omega^2\tilde{p}(\omega)-i\omega\gamma_b\tilde{p}(\omega)+\omega_b^2\tilde{p}(\omega)+\kappa_1\tilde{q}(\omega)=-(e^2/m)\tilde{E}_{inc}(\omega)$$
$$-\omega^2\tilde{q}(\omega)-i\omega\gamma_d\tilde{q}(\omega)+\omega_d^2\tilde{q}(\omega)+\kappa_2\tilde{p}(\omega)=-\omega^2M\tilde{p}(\omega) \quad (2)$$

Where the damping coefficients of dipole and quadruple mode are $\gamma_b = R_1/L_1$ and $\gamma_d = R_2/L_2$, respectively, and their resonant frequency $\omega_b$ and $\omega_d$ follow as:

$$\omega_b = 1/\sqrt{L_1 C_b},\quad \omega_d = 1/\sqrt{L_2 C_d} \quad (3)$$

The capacity $C_b$ and $C_d$ in each partial circuit follow as:

$$\frac{1}{C_b}=\frac{1}{C_1}+\frac{1}{C_c},\quad \frac{1}{C_d}=\frac{1}{C_2}+\frac{1}{C_c} \quad (4)$$

Contrary to previous paper, the coupling constant $\kappa_1$ and $\kappa_2$ are different from each other, follow as:

$$\kappa_1 = \frac{1}{L_1 C_c},\quad \kappa_2 = \frac{1}{L_2 C_c} \quad (5)$$

The essence of these coupling will be developed at later paper, these depend on the parallel connection of bright mode and dark mode' capacitors and their strength are inverse proportion to the distance between bright mode and dark mode.

In above equation, if mutual induction M is very small as in the dolmen structure, or there is no such term as in the T shape structure, it has the following typical two coupled oscillator model[1,43].

$$-\omega^2\tilde{p}(\omega)-i\omega\gamma_b\tilde{p}(\omega)+\omega_b^2\tilde{p}(\omega)+\kappa_1\tilde{q}(\omega)=-(e^2/m)\tilde{E}_{inc}(\omega)$$
$$-\omega^2\tilde{q}(\omega)-i\omega\gamma_d\tilde{q}(\omega)+\omega_d^2\tilde{q}(\omega)+\kappa_2\tilde{p}(\omega)=0 \quad (6)$$

Where $\tilde{E}_{inc}(\omega)$ is the strength of the external electrical field and $E(t)=\tilde{E}_{inc}e^{-i\omega t}$. Of course, if one take account of the mutual induction, the coefficient of the radiative oscillator equation vary a bit from, but they have the same formula and their solutions is.

$$\tilde{p}(\omega)=(e^2/m)\frac{D_d(\omega)\tilde{E}_{inc}(\omega)}{D_d(\omega)D_b(\omega)-\kappa_1\kappa_2}$$
$$\tilde{q}(\omega)=(e^2/m)\frac{\kappa_2\tilde{E}_{inc}(\omega)}{D_d(\omega)D_b(\omega)-\kappa_1\kappa_2} \quad (7)$$

Where $D_b(\omega)$ and $D_d(\omega)$ is

$$D_b(\omega)=\omega_b^2-i\omega\gamma_b-\omega^2$$
$$D_d(\omega)=\omega_d^2-i\omega\gamma_d-\omega^2 \quad (8)$$

Now taking the following quantity,

$$\chi_0 = \frac{ne^2}{\varepsilon_0 m\omega_b^2}=\frac{\omega_p^2}{\omega_b^2} \quad (9),$$

then the total average polarization of dipole modes has the following equation.

$$\tilde{P}(\omega)=\varepsilon_0\chi(\omega)\tilde{E}_s(\omega)=n\delta(\omega)\tilde{p}(\omega) \quad (10)$$

Where $\omega_p$ is the plasma frequency of metal, $\tilde{E}_s(\omega)$ is the surface field, $n$ is the atom number per unit of volume, $\delta(\omega)$ is the quantity as much as penetration depth of the external field, $n\delta(\omega)$ is the atom number per unit of surface area[1].

$\chi(\omega)$ is the frequency dependent susceptibility at meta-surface, has the following form.

$$\chi(\omega) = \chi_0 \frac{D_d(\omega)\omega_b^2}{D_d(\omega)D_b(\omega) - \kappa_1\kappa_2} \tag{11}$$

From (10), the relation between macro-surface electrical field and the external field is

$$\tilde{E}_s(\omega) = \delta(\omega)\tilde{E}_{inc}(\omega). \tag{12}$$

The total average polarization thus equals $\tilde{P}(\omega) = \varepsilon_0\chi(\omega)\delta(\omega)\tilde{E}_{inc}(\omega)$ and the polarization current at meta-surface is $\langle \tilde{j}_s(\omega)\rangle = -i\omega\tilde{P}(\omega)$ in the harmonic time approach.

From the relationship current and electrical field, $\langle \tilde{j}_s(\omega)\rangle = \sigma_{se}\tilde{E}_s(\omega)$, we can derive the equation on surface conductivity,

$$\sigma_{se} = -\frac{i\omega P(\omega)}{\tilde{E}_s(\omega)} = -\frac{i\varepsilon_0\omega\delta(\omega)\omega_b^2 D_d(\omega)}{D_d(\omega)D_b(\omega) - \kappa_1\kappa_2} \tag{13}$$

Define the following the quantity,

$$S = \zeta\sigma_{se} = -\frac{i\eta\omega_b^2 D_d(\omega)}{D_d(\omega)D_b(\omega) - \kappa_1\kappa_2} \quad . \tag{14}$$

Where $\zeta$ is the impedance of the external electromagnetic wave, that is $\zeta = \mu_0\omega/k$, $\mu_0$ is the permeability, $k$ is the wave number of the external electromagnetic wave. And $\eta$ is $\eta = \varepsilon_0\mu_0\delta(\omega)\omega^2/k$, in general, is the dimensionless quantity as much as one.

Following the method of [1], in the system comprised bright and dark mode excited by external electromagnetic field, the scattering parameters of the planar metamaterials are

$$r = -\frac{S}{2+S}, \quad t = \frac{2}{2+S} \quad . \tag{15}$$

Following the formulism of [4, 7] using S defined from (15), derive the Fano-like formula on the reflectance, transmittance and absorption near the bright and dark resonances of planar metamaterials.

First, derive the formula on the reflectance near dark resonance,

$$R = |r|^2 = \left|\frac{S}{2+S}\right|^2 = \frac{\omega_b^4|\eta|^2}{4}\left|\frac{D_d(\omega)}{D_d(\omega)D_{rb}(\omega) - \kappa_1\kappa_2}\right|^2 \quad . \tag{16}$$

Where $D_{rb}(\omega)$ is the following strength of modified radiative oscillator,

$$D_{rb}(\omega) = \omega_b^2 - i(\omega\gamma_b + \eta\omega_b^2/2) - \omega^2 = \omega_b^2 - i\omega\gamma_{rb} - \omega^2 \quad . \tag{17}$$

From (17), the reflectance can be written as

$$R = \frac{\omega_b^4|\eta|^2}{4|D_{rb}(\omega)|^2}\left|\frac{D_d(\omega)}{D_d(\omega) - \kappa_1\kappa_2\,\text{Re}(D_{rb}(\omega))/|D_{rb}(\omega)|^2 - i\kappa_1\kappa_2\,\text{Im}(D_{rb}(\omega))/|D_{rb}(\omega)|^2}\right|^2$$

Where $\Delta_{rd} = -\kappa_1\kappa_2\,\text{Re}(D_{rb}(\omega_d))/\omega_d|D_{rb}(\omega_d)|^2$ is the shift resonance position of the dark mode by the resonant coupling with the bright mode, $\Gamma_d = \gamma_d \omega_d$ is the resonant width by the intrinsic loss of the dark mode, $\Gamma_{rdc} = \kappa_1 \kappa_2 \, \text{Im}(D_{rb}(\omega_d))/|D_{rb}(\omega_d)|^2$ is the radiative resonant width occurring by the coupling the dark mode with bright mode and $\Gamma_{Trd} = \Gamma_d + \Gamma_{drc}$ is the total resonant width.

Then the reflectance near the natural frequency of the dark mode, $R_d$ is given by:

$$R_d = \frac{\omega_b^4 |\eta|^2}{4|D_{rb}|^2} \left| \frac{-\text{Re}\,D_d + i\Gamma_d}{-\text{Re}\,D_d + i\Gamma_d - \omega_d \Delta_{rd} + i\Gamma_{rdc}} \right|^2 = \frac{\omega_b^4 |\eta|^2}{4|D_{rb}|^2} \left| \frac{-\text{Re}\,D_d + i\Gamma_d}{-\text{Re}\,D_d - \omega_d \Delta_{rd} + i\Gamma_{Trd}} \right|^2$$

Now, define the following quantities.

$$\varepsilon_{rd} = \frac{\omega^2 - \omega_d^2 - \omega_d \Delta_{rd}}{\Gamma_{Trd}}, \tag{18}$$

$\varepsilon_{rd}$ is the reduced frequency at the shifted resonant position, that is, the Fano resonant frequency, and

$$q_{rd} = -\frac{\text{Re}\,D_{rd}(\omega_d)}{\text{Im}\,D_{rd}(\omega_d)(1 + \Gamma_d/\Gamma_{rdc})} = \frac{\omega_d \Delta_{rd}}{\Gamma_{Trd}}, \tag{19}$$

$q_{rd}$ is the asymmetric factor of the spectrum near the Fano resonant frequency of the dark mode, then the reflectance near the Fano resonant frequency of the dark mode has the following the Fano-like spectrum.

$$R_d = \frac{\omega_b^4 |\eta|^2}{4|D_{rb}(\omega_d)|^2} \frac{(\varepsilon_{rd} + q_{rd})^2 + b_{rd}}{\varepsilon_{rd}^2 + 1} \tag{20}$$

Where $b_{rd} = \Gamma_d^2/\Gamma_{Trd}^2$ is the modulation damping parameter presenting the damping rate of the non-radiative loss that prevent full interference to the total loss [4],

Then, consider transmittance near the Fano resonant frequency of the dark mode using S.

$$T = |t|^2 = \left|\frac{2}{2+S}\right|^2 = \left|\frac{D_d(\omega)D_b(\omega) - \kappa_1\kappa_2}{D_d(\omega)D_{rb}(\omega) - \kappa_1\kappa_2}\right|^2$$

$$= \frac{|D_b|^2}{|D_{rb}|^2} \left| \frac{D_d}{-D_d + \kappa_1\kappa_2 \, \text{Re}\,D_{rb}/|D_{rb}|^2 + i\kappa_1\kappa_2 \, \text{Im}\,D_{rb}/|D_{rb}|^2} \right|^2 \times$$

$$\times \left| \frac{-D_d + \kappa_1\kappa_2 \, \text{Re}\,D_b/|D_b|^2 + i\kappa_1\kappa_2 \, \text{Im}\,D_b/|D_b|^2}{D_d} \right|^2$$

Accepting new reduced frequency, $\varepsilon_d = \omega^2 - \omega_d^2 - \omega_d \Delta_d/\Gamma_{Td}$, the resonance frequency shift $\Delta_d = \kappa_1\kappa_2 \, \text{Re}\,D_b(\omega_d)/\omega_d|D_b(\omega_d)|^2$ and the asymmetric parameter $q_d = \omega_d \Delta_d/\Gamma_{Td}$,

$$T_d = \frac{|D_b(\omega_d)|^2}{|D_{rb}(\omega_d)|^2} \frac{(\varepsilon_{rd} + q_{rd})^2 + b_{rd}}{\varepsilon_{rd}^2 + 1} \frac{\varepsilon_d^2 + 1}{(\varepsilon_d + q_d)^2 + b_d} \tag{21}$$

Where $\Gamma_{dc}$ is defined as $\Gamma_{dc} = \kappa_1\kappa_2 \, \text{Im}\,D_b(\omega_d)/|D_b(\omega_d)|^2$, $\Gamma_{Td} = \Gamma_d + \Gamma_{dc}$ is new resonance width and $b_d = \Gamma_d^2/\Gamma_{Td}^2$ is new modulation damping parameter.

Now, transform $\varepsilon_d$, $q_d$, $b_d$ into the following quantities,

$$\varepsilon_{md} = (\varepsilon_d + q_d)/\sqrt{b_d} \tag{22a}$$

$$q_{md} = -q_d/\sqrt{b_d} \tag{22b}$$

$$b_{md} = 1/b_d, \tag{22c}$$

the formula of the transmittance then transforms into the product of the the Fano formula.

$$T_d = \frac{|D_b(\omega_d)|^2}{|D_{rb}(\omega_d)|^2} \frac{(\varepsilon_{rd} + q_{rd})^2 + b_{rd}}{\varepsilon_{rd}^2 + 1} \frac{(\varepsilon_{md} + q_{md})^2 + b_{md}}{\varepsilon_{md}^2 + 1} \tag{23}$$

To consider the shapes of reflectance and transmittance near the resonance position of the bright mode, we transform the reflectance into the following form.

$$R = |r|^2 = \left|\frac{S}{2+S}\right|^2 = \frac{\omega_b^4 |\eta|^2}{4} \left|\frac{D_d(\omega)}{D_d(\omega)D_{rb}(\omega) - \kappa_1\kappa_2}\right|^2$$

$$= \frac{\omega_b^4 |\eta|^2}{4} \left|\frac{1}{D_{rb}(\omega) - \kappa_1\kappa_2/D_d(\omega)}\right|^2$$

$$= \frac{\omega_b^4 |\eta|^2}{4} \left|\frac{1}{D_{rb}(\omega) - \kappa_1\kappa_2 \operatorname{Re} D_d(\omega)/|D_d(\omega)|^2 + i\kappa_1\kappa_2 \operatorname{Im} D_d(\omega)/|D_d(\omega)|^2}\right|^2$$

We define the radiative resonance of bright mode by coupling with the dark mode as $\Gamma_{bc} = \kappa_1\kappa_2 \operatorname{Im} D_d(\omega_b)/|D_d(\omega_b)|^2$, then the total resonant width near the natural frequency of the bright mode is $\Gamma_{Trb} = \Gamma_{bc} - \operatorname{Im} D_{rb}(\omega_b) = \Gamma_{bc} + \omega_b\gamma_b + \eta\omega_b^2/2$. Also, the reduced frequency and the shift from the natural frequency of the dipole mode is respectively defined by:

$$\varepsilon_{rb} = \frac{\omega^2 - \omega_b^2 - \omega_b\Delta_b}{\Gamma_{Trb}} \tag{24}$$

$$\Delta_b = -\frac{\kappa_1\kappa_2 \operatorname{Re} D_d(\omega_b)}{\omega_b |D_d(\omega_b)|^2} \tag{25}$$

The line-shape of the reflectance near the natural frequency of the bright mode is given by:

$$R_b = \frac{\omega_b^4 |\eta|^2}{4} \left|\frac{1}{-D_{rb}(\omega) - \omega_b\Delta_b + i\Gamma_{bc}}\right|^2 = \frac{\omega_b^4 |\eta|^2}{4\Gamma_{Trb}^2} \frac{1}{\varepsilon_{rb}^2 + 1} \tag{26}$$

Thus, the reflectance near the natural frequency of the bright mode has the symmetric shape, that is, not the Fano form but Lorentzian.

Next, the transmittance near the natural frequency of the bright mode is derived and transformed into the following form, similarly to the dark mode'.

$$T = |t|^2 = \left|\frac{2}{2+S}\right|^2 = \left|\frac{D_d(\omega)D_b(\omega) - \kappa_1\kappa_2}{D_d(\omega)D_{rb}(\omega) - \kappa_1\kappa_2}\right|^2$$

Now near the natural frequency of the bright mode, the reduced frequency is defined by:

$$\varepsilon_b = \omega^2 - \omega_b^2 - \omega_b \Delta_b / \Gamma_{Tb},$$

The total resonant width is defined by:

$$\Gamma_{Tb} = \Gamma_{bc} - \operatorname{Im} D_b(\omega_b) = \Gamma_{bc} + \omega_b \gamma_b$$

And the modulation damping parameter is defined by:

$$b_b = |\Gamma_{Tb}/\Gamma_{Trb}|^2,$$

the transmittance near the natural frequency of the bright mode is given by:

$$T_b = \left| \frac{-D_b(\omega) - \Delta_b \omega_b + i\Gamma_{bc}}{-D_{rb}(\omega) - \Delta_b \omega_b + i\Gamma_{bc}} \right|^2 = \left| \frac{\Gamma_{Tb}}{\Gamma_{Trb}} \right|^2 \frac{\varepsilon_b^2 + 1}{\varepsilon_{rb}^2 + 1} = \frac{\varepsilon_{rb}^2 + b_b}{\varepsilon_{rb}^2 + 1}. \tag{27}$$

From these formulas, it is found that both the shifts near the natural frequency of the bright mode are equal to, either the reflectance and transmittance exhibit the symmetric spectrum.

Finally, consider the absorption near the natural frequency of the bright mode.

The absorption is given as the following form by [1]:

$$A = T \operatorname{Re} S \tag{28}$$

Here,

$$\operatorname{Re} S = \operatorname{Re}\left( \frac{i\eta\omega_b^2 D_d(\omega)}{D_d(\omega)D_d(\omega) - \kappa_1\kappa_2} \right)$$

$$= \frac{\eta\omega_b^2}{\Gamma_{Tb}} \operatorname{Re}\left( \frac{i}{\varepsilon_b + i} \right) = \frac{\eta\omega_b^2}{\Gamma_{Tb}(\varepsilon_b^2 + 1)}$$

Coupling this formula with the formula (17) on the transmittance, the absorption A near the resonant frequency is given by:

$$A = \eta\omega_b^2 \frac{\Gamma_{Tb}}{\Gamma_{Trb}^2} \frac{1}{\varepsilon_{rb}^2 + 1} \tag{29}$$

This formula is very importance when we study the electromagnetically induced transparency and absorption. With this formula, we can obtain the condition to design the electromagnetically transparency and absorption in metamaterials.

## 2. Results and discussion

The signification of the Fano formula is simply to explain the complicated characters of the spectrum appearing in the system comprised of the continuum and discrete states. in terms of three factors, that is, the reduced energy $\varepsilon$ defined by $\varepsilon = 2(E - E_0)/\Gamma$ (or reduced frequency defined by $2(\omega - \omega_0)/\Gamma$, where $\omega_0$ is the resonance position, $\Gamma$ is the resonant width), the asymmetric shape factor q and the modulation damping parameter b.[4,7,38]

Many optical spectrums appearing in the metamaterials and nano-plasmonics have the Fano profile, their mechanism can be explained by two coupled harmonic oscillators.

One of the applications in metamaterials and nano-plasmonics is the electro-optic control of light due to the interaction of bright and dark mode. In general, the optical properties occurring by the interaction of bright and dark mode are determined in terms of the geometry of bright and dark mode, and the their configuration. The geometry of bright mode and dark mode determines the natural frequencies of these modes and their configuration determines the coupling constants. The optical properties occurring due to the configuration of bright and dark mode discuss in [4, 42].

In this paper, we discuss the optical properties due to the difference of the natural frequencies between bright and dark mode, that is, detuning.

First, consider (21) and (27) on the reflectance near the natural frequency of bright mode and dark mode. From (21), the line-shape of reflectance near the natural frequency of dark mode has the asymmetric character, thus exhibits the Fano form affected by the bright mode. Also the reflectance near the natural frequency of bright mode has the symmetric shape, that is, Lorenzian from (27)

The spectrum shape near the natural frequency of bright mode is almost determined by the character of bright mode and it means that the interaction with the dark mode near the natural frequency of bright mode is weak. As shown in the reflectance of Figure1, the shape of reflectance is similar to the one of bright mode in the case of no coupling except near the natural frequency of the dark mode.

As the interaction increases, the shape of the reflectance near the resonance of dark mode is governed by the Fano form, as the spectral detuning between the bright and dark resonances is close to zero, it approach to symmetry and has a dip in the reflection spectrum, it thus becomes the anti-resonance.
This means that the destructive interference occurs between the direct excitation of the bright mode and the external field and the indirect excitation of the dark mode, the classical analog of the electromagnetically induced transparency arises in the vicinity of the dark resonance [9].

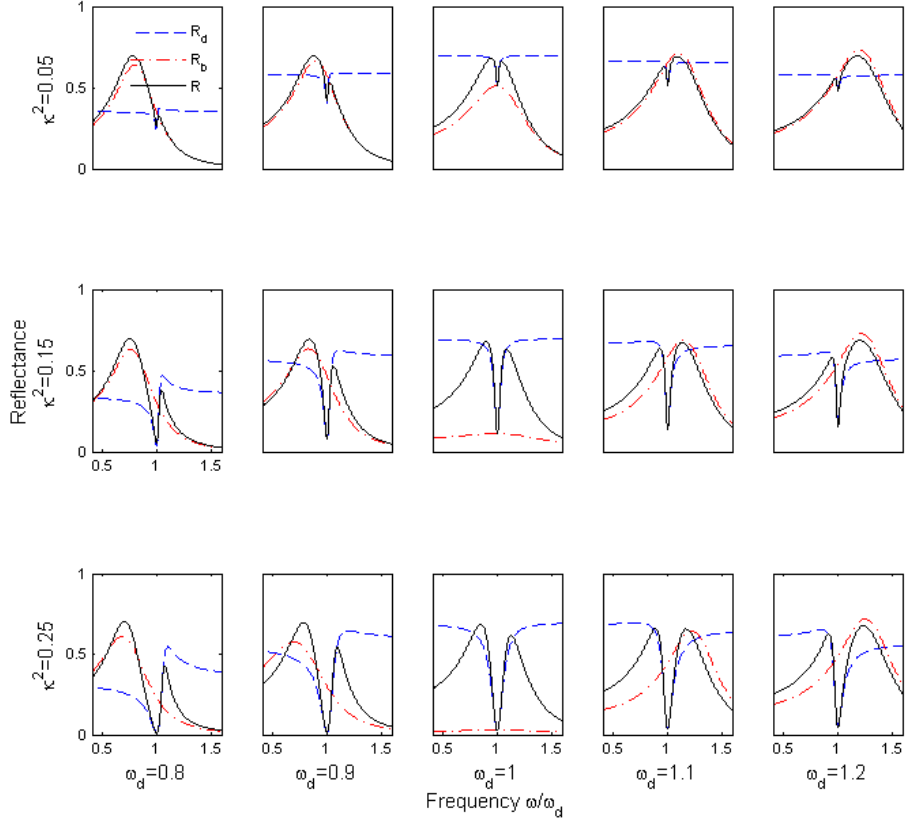

Figure 1. The reflectance via the detuning between the resonances of the bright and dark mode. From top to down, the coupling increase. (Black solid line) full reflectance, (blue dashed line) the reflectance of (21) and (red dot-dashed line) the reflectance of (21)

Figue 2 shows how the resonance position of the dark mode is shifted in terms of the interference of bright and dark modes. The resonant frequency often call the Fano resonant frequency, is given by (21).

$$\omega_0 \approx \omega_d + \Delta/2 \qquad (30)$$

Where $\omega_d$ is the natural frequency when is no coupling.

The resonance position of the dark mode is related to the interaction between the bright and dark modes, so that the stronger the interaction, the father is shifted its resonant position. In what direction the resonance position is shifted is relate to the relative difference between the resonance positions of the bright and dark mode, and in case that the natural frequency of a dark mode is smaller than the one of a bright mode, the resonance of dark mode is blue-shifted and in opposite case, the resonance of dark mode is red-shifted.

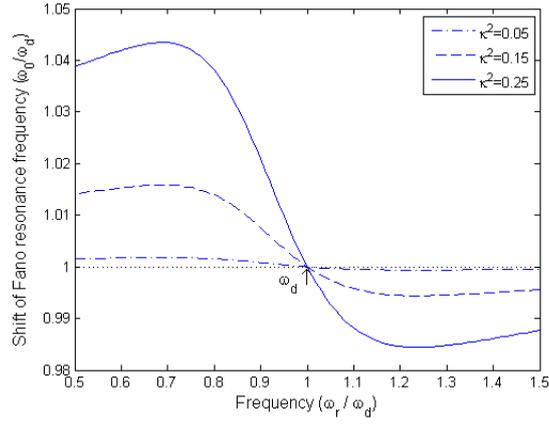

Figure 2. the shifting of the dark resonance via the intentsity.

In general , the shape of the reflection spectrum result in the standard Lorenzian profile as $|q| \to \infty$, the total shape of the reflection spectrum thus appears the symmetric one having a pick. Then the interaction with the bright mode is very weak, so that it result in the resonant feature of the no coupling case. When $q$ is zero, the shape of the reflection spectrum becomes anti-resonant state and the total shape of reflection spectrum appears as the symmetric one having a dip []. When $q$ is finite, an asymmetric shape is obtained in the reflection spectrum, according to the value of q in the reference of $q=1$, we can discuss the influence of the bright and dark modes.

As shown figure 3, the asymmetric shape factor $q_{rd}$ of reflectance of the dark mode usually has a finite value when the detuning between the bright and dark modes is finite, the smaller the detuning, the closer approach zero, so that it becomes anti-resonant state. This means that the reflectance and transmittance leads to typical Fano form in the vicinity of the dark resonance, when the detuning of the bright and dark resonances, the special state of Fano interference, that is, this leads to the classical analog of electromagnetically induced transparency or absorption [38,43].

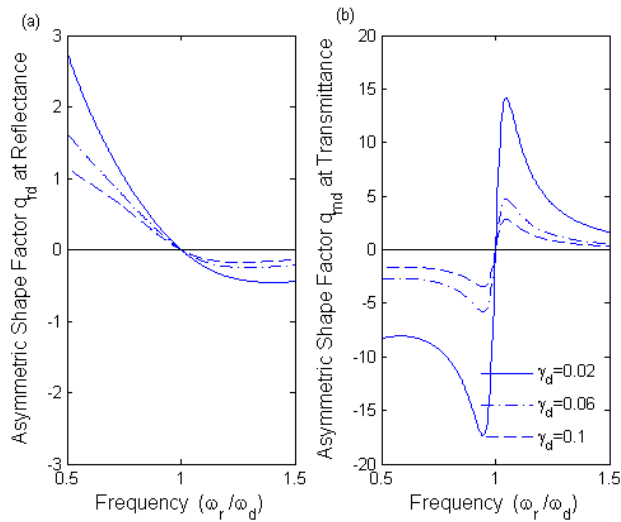

Figure 3. The asymmetric shape factor $q_{rd}$ (a) and $q_{md}$ (b) of Fano form via the magnitude of the intrinsic loss of the dark mode in the vicinity of dark resonance.

Secondly, discuss the transmittance of the bright and dark modes.

Main feature is the shape of the formula on the transmittance near dark resonance. The eq (23) on the transmittance is comprised of the product of the Fano form equal to reflectance and another Fano form. This new Fano form plays an important role in the classical analog of electromagnetically in deduced transparency.

As shown figure 3, the asymmetric shape factor of new Fano form usually has a finite value when the detuning between the natural frequencies is finite, it abruptly increases as the detuning decreases and thus becomes the resonant Lorentzian.

Next, consider the transmittance near the natural frequency of the bright mode. As shown eq (27) and the red dot-dashed curve of figure 4, the transmittance has a dip and is symmetric in the vicinity of the natural frequency of the bright mode. This corresponds to $q=0$ in the Fano formula and shows the anti-resonant state. Hence when the detuning is large, main feature of the transmittance near the natural frequency of the bright mode is mostly determined by the bright mode.

When the detuning is zero, the maximum of the transmittance of the dark mode and the minimum of the transmittance coincide with a point and this is a special type of electromagnetically induced transparency and absorption. As know above discussion, the interference between the bright and dark modes mostly occurs in vicinity of the Fano resonance position, that is, ep (30). In order to consider in detail the effect of the interference in the vicinity of Fano resonant point, calculate the rate of the radiative loss to the intrinsic loss.

If the radiative loss is large, this means that the interference between the bright and dark modes is strong and from the viewpoint of the dark mode, indirect interaction with the external field take actively place due to the interference with the bright mode. According to the Fano formulism, this means that the destructive or constructive interference take place between direct interference pathway with the external field and the indirect one. When the destructive interference take place, it result in electromagnetically induced transparency and when the constructive interference take place, it result in electromagnetically induced absorption.

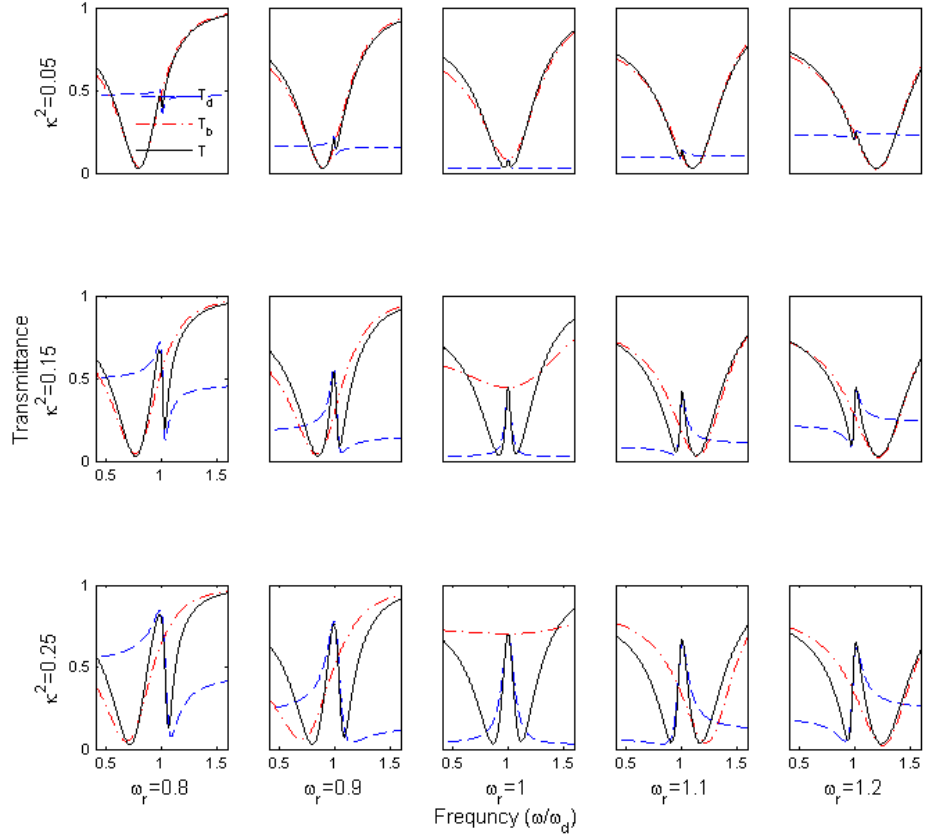

Figure 4. The transmittance via the detuning between the natural frequencies of the bright mode and dark mode. From top to down, the coupling increase. (Black solid line) full transmittance, (blue dashed line) the transmittance of eq (23) and (red dot-dashed line) the transmittance of eq (27)

Figure 5 shows the rate of the radiative loss to the intrinsic loss of bright and dark modes. Hence, we know that the interference of the dark mode exists in the broad detuning interval and in contrast to this, the interference of the bright mode take place in the case that the detuning is relatively small due to the coupling. This means that the interference of the bright mode with the external field or the interference of the bright mode with the dark mode take strongly place when the detuning is small.

Finally, consider the energy stored in the dark mode according to the detuning of the bright and dark mode. This energy stored in the dark mode relates to the modulation damping factor [4]. From calculation, (1) the stronger the coupling between the bright mode and dark mode, (2) the smaller the intrinsic loss of the dark mode, the more energy is stored in the dark mode. We know that the most energy can be stored in the dark mode when the detuning becomes zero, that is, when the natural frequencies of the bright and the dark modes coincide.

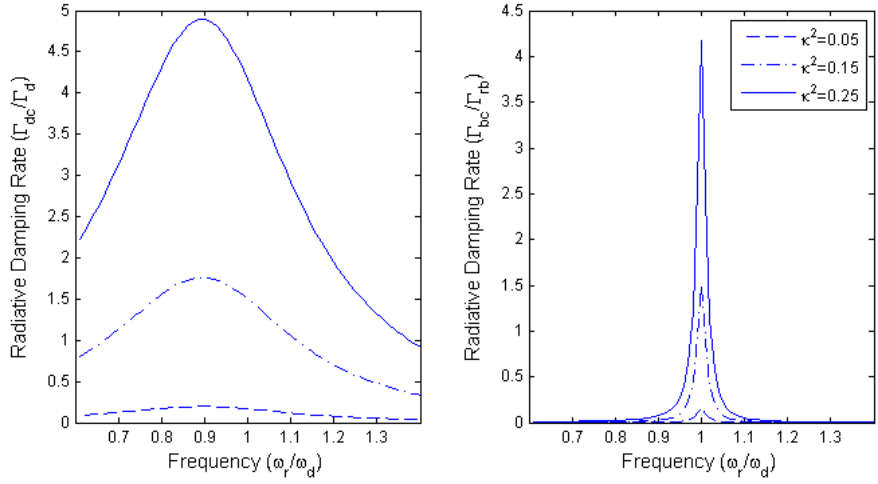

Figure 5. The radiative loss rate of the bright and dark modes

## Conclusion

In this paper, we discuss the effect of the bright and dark modes on the optical spectrum in planar metamaterials, based on the two coupled oscillator model.

We derive the Fano formulism of the optical spectrum in the vicinity of the dark and bright resonances and analysis the effect of the individual factors on the total spectrum. This formulism is very useful to analysis the electromagnetically induced transparency and absorption and can give the important key to design the optical device in metamaterials and the plasmonics.